# Dynamic manipulation in piezoresponse force microscopy: creating non-equilibrium phases with large electromechanical response

Kyle P. Kelley,[1] Yao Ren,[2] Anna N. Morozovska,[3] Eugene A. Eliseev,[4] Yoshitaka Ehara,[5,6] Hiroshi Funakubo,[6] Thierry Giamarchi,[7] Nina Balke,[1] Rama K. Vasudevan,[1] Ye Cao,[2] Stephen Jesse,[1] and Sergei V. Kalinin[1]

[1] Center for Nanophase Materials Sciences, Oak Ridge National Laboratory, Oak Ridge, TN 37831

[2] Department of Materials Science and Engineering, University of Texas at Arlington, Arlington, TX 76019

[3] Institute of Physics, National Academy of Science of Ukraine, pr. Nauki 46, 03028 Kyiv, Ukraine

[4] Institute for Problems of Materials Science, National Academy of Science of Ukraine, Krjijanovskogo 3, 03142 Kyiv, Ukraine,

[5] Department of Communications Engineering, National Defense Academy, Hashirimizu, Yokosuka, 239-8686, Japan.

[6] Department of Material Science and Engineering, Tokyo Institute of Technology, Yokohama 226-8502, Japan

[7] Department of Quantum Matter Physics, University of Geneva, 24 Quai Ernest-Ansermet, CH-1211 Geneva, Switzerland

Domains walls and topological defects in ferroelectric materials have emerged as a powerful new paradigm for functional electronic devices including memory and logic. Similarly, wall interactions and dynamics underpin a broad range of mesoscale phenomena ranging from giant electromechanical responses to memory effects. Exploring the functionalities of individual domain walls, their interactions, and controlled modifications of the domain structures is crucial for applications and fundamental physical studies. However, the dynamic nature of these features severely limits studies of their local physics since application of local biases or pressures in piezoresponse force microscopy induce wall displacement as a primary response. Here, we introduce a fundamentally new approach for the control and modification of domain structures based on automated experimentation whereby real space image-based feedback is used to control the tip bias during ferroelectric switching, allowing for modification routes conditioned on domain states under the tip. This automated experiment approach is demonstrated for the exploration of domain wall dynamics and creation of metastable phases with large electromechanical response.

Since their discovery in the beginning of 20th century, ferroelectric materials remain one of the most fascinating areas in the field of condensed matter physics. The basic physics of these materials stem from the presence of unit-cell level dipoles ordered on the macroscopic length scales giving rise to a number of unique functionalities ranging from high dielectric constants to electromechanical coupling and electrooptical properties. When combined with the electrical field control of polarization, this lays the foundation for a broad spectrum of applications ranging from MEMS[1–3] and NEMS[4,5] to electrooptical devices[6,7] to information technology devices such as ferroelectric field effect transistors,[8–11] tunneling barriers,[12–15] and domain wall microelectronics.[16–19]

However, an even broader spectrum of functionalities emerge on the mesoscopic scale due to the interplay between topological defects, discontinuities in the polarization fields, frozen structures, and charge disorder in the material mediated by long-range electrostatic and mechanical depolarization fields. A well-recognized example of such behavior is the enhanced electromechanical response in ferroelectric ceramics associated with non-180° domain wall motions. These behaviors are most pronounced in ferroelectric relaxors and morphotropic phase boundary materials, in which the presence of a large density of topological defects induced by local disorder and domain splitting in the proximity of symmetry incompatible phases gives rise to the giant dielectric and electromechanical responses.

In all these cases, the mechanisms behind these behaviors are stochastic in nature and the phases with unique responses emerge as a result of the interplay between disorder and symmetry incompatible phase competition. A number of attempts have been made to harness these behaviors and create materials with enhanced properties by design. One such example can be directional poling in non-major crystallographic directions with subsequent annealing. Another example of such control is the careful tuning of the material microstructure in multilayers (for example, compositional gradients or superlattice structures with dielectric spacers). Here, balancing the characteristic size of ferroelectric regions and controlling depolarization effects via the interface layers enables a broad set of domain structures including topological vortex states.[20]

Global methods for materials control are necessarily limited. In parallel, much attention has been paid to the local control of ferroelectric domain structures via piezoresponse force microscopy (PFM). The classical, single-frequency PFM and its band-excitation and G-mode variants have emerged as primary techniques for the visualization of ferroelectric domains on the nanometer scale. At the same time, a biased PFM tip can be used to switch normal and, under certain conditions, in-plane components of polarization. While in most cases PFM polarization switching is used to establish the presence of switchable polarization, a number of groups have used PFM to create predetermined domain structures and topological defects and to explore their functional properties such as conductivity and microwave responses.[16,18,19,21–27] These advances set the foundation for emergent areas such as domain wall oxide electronics, resulting in continuous interest in domain wall properties and functionalities over the last decade.

Yet, for over 25 years since the invention of PFM, tip-induced manipulation has invariably been based on the application of predetermined bias pulses and directional poling (Fig. 1a). This approach relies on the known relationship between the applied bias history and resulting domain structure. In this case, the tip bias is tuned to provide the desired domain structure and materials with sufficiently high wall pinning and low frozen disorder are used to avoid spontaneous wall roughening, back switching, and other deleterious effects. This approach can be further combined with tip-field symmetry breaking via lateral probe motion to enable non-180° switching.[28] However, domain structures and hence domain wall topology are ascertained only before and after the writing experiment, while domain structure evolution during the poling process is unknown. Similarly, this approach cannot be used to study polarization dynamics of objects such as 2D ferroelectric domain walls or 1D topological defects, such as wall junctions or vortices, that can easily move in the surface plane.[29–31] Broad areas of ferroic physics of direct interest, for both applications (domain wall conduction[23–26,32–34] and topological defects[32,33,35,36]) and fundamental studies (pinning mechanisms[37], etc.) have been explored, but details on the precise dynamics are often difficult to extract given these experimental limitations.

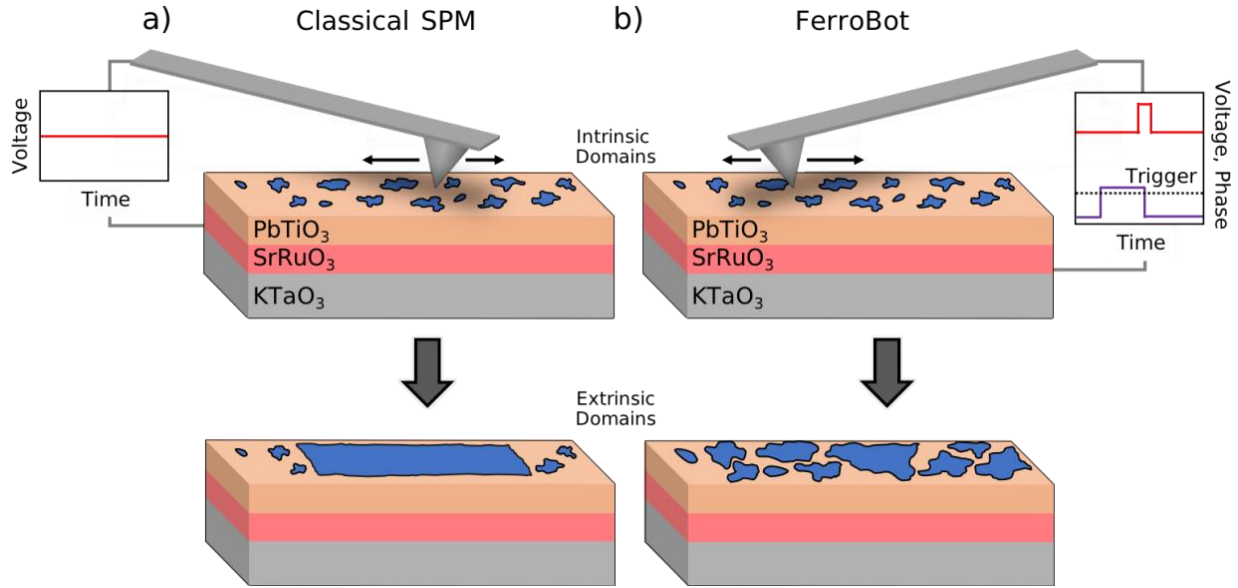

**Figure 1.** Existing paradigms for ferroelectric domain switching. (a) In classical SPM, poling is based on predefined pulse sequence, or constant bias, to create the pattern. (b) In FerroBot, the PFM signal is detected by an image-based feedback circuit and writing sequence is generated based on detected signal. In the simple implementation of FerroBot, a sequence of switching pulses with given magnitude, duration, and delay is generated upon crossing the ferroelectric domain wall. Note, the pulse sequence can be conditioned based on the wall polarity, motion direction, etc., exploring complex conditional sequences.

Here, we demonstrate a fundamentally new approach for exploring polarization dynamics of specific domain elements and topological structures based on *in-situ* image-based feedback in PFM. This automated experiment approach is demonstrated for probing highly pinned domain wall dynamics in polydomain $PbTiO_3$ (PTO), a traditionally difficult subject for PFM studies. We further demonstrate the creation of frustrated phases with a significantly enhanced electromechanical response prior to the onset of wall motion. These studies provide a pathway for novel applications of PFM and more generally for automated experiments in scanning probe microscopy (SPM) for exploring physics of low dimensional systems and material control.

The central concept of the automated PFM experiment, the FerroBot, is illustrated in Figure 1b. The main set-up is similar to classical PFM and can be implemented in single frequency, band excitation[38,39], or G-modes[40,41]. To enable the automated experiment approach, the detected signal during PFM scans, e.g., the amplitude and phase of the electromechanical response for single-frequency detection, is set to the logical circuit that detects the specific local event, such as crossing the domain wall of a defined type, detection of a specific feature, or a more complex descriptor. Upon detection of the event, the FerroBot can be configured to produce a user-defined set of actions, for example, generating a pulse with a specific delay time and magnitude dependent on the stimulus (Fig. 1b). We note that the unique advantage of the proposed approach is that the feedback is fully logical, and hence, the performed action can be predicated on the specific type of event, e.g., crossing one or several walls, arbitrary delay times, etc. This action can also be predicated on global control variables such as the type of the local domain structure. However, here we focus on the simplest implementation based on single wall detection.

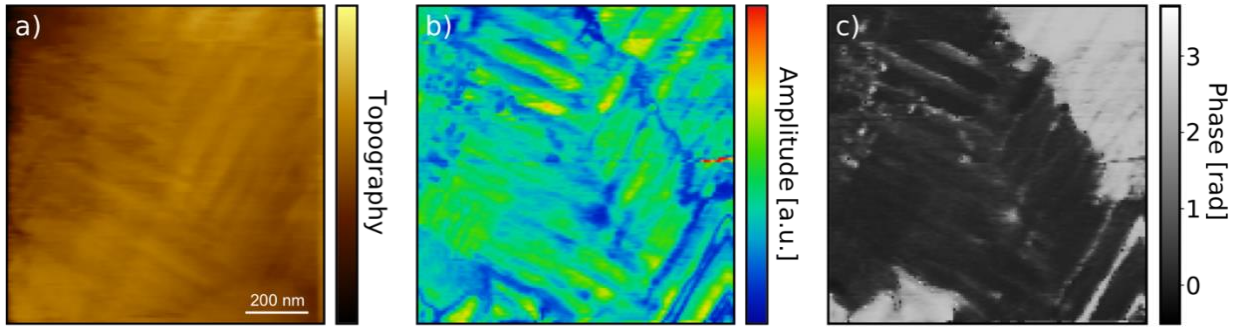

**Figure 2:** a) Surface topography of 700 nm PTO film on $SrRuO_3$/$KTiO_3$ measured via AFM, and single frequency PFM; b) amplitude and c) phase illustrating dense *a-c* domain structure.

Here, the experiments were performed on an ultra-high vacuum Omicron atomic force microscope/scanning tunneling microscope (AFM-STM) modified for the PFM measurements as described elsewhere.[42] The FerroBot uses the single-frequency PFM phase as a detector for the 180° domain walls where the wall crossing event is used as a trigger for variable voltage pulses, chosen to be 10 ms here. The trigger for FerroBot was implemented from a user defined input phase threshold, where once the phase changes by a given amount, a voltage pulse is applied to the AFM cantilever allowing for quantitative characterization of domain wall motion. Further experimental details are provided in the methods section. As a model system, we have chosen polydomain 700nm PTO grown on $KTiO_3$ with a $SrRuO_3$ back electrode. As shown in Figure 2, the system possesses a dense *a-c* domain structure, providing an ideal platform for understanding domain wall motion in a highly pinned system.

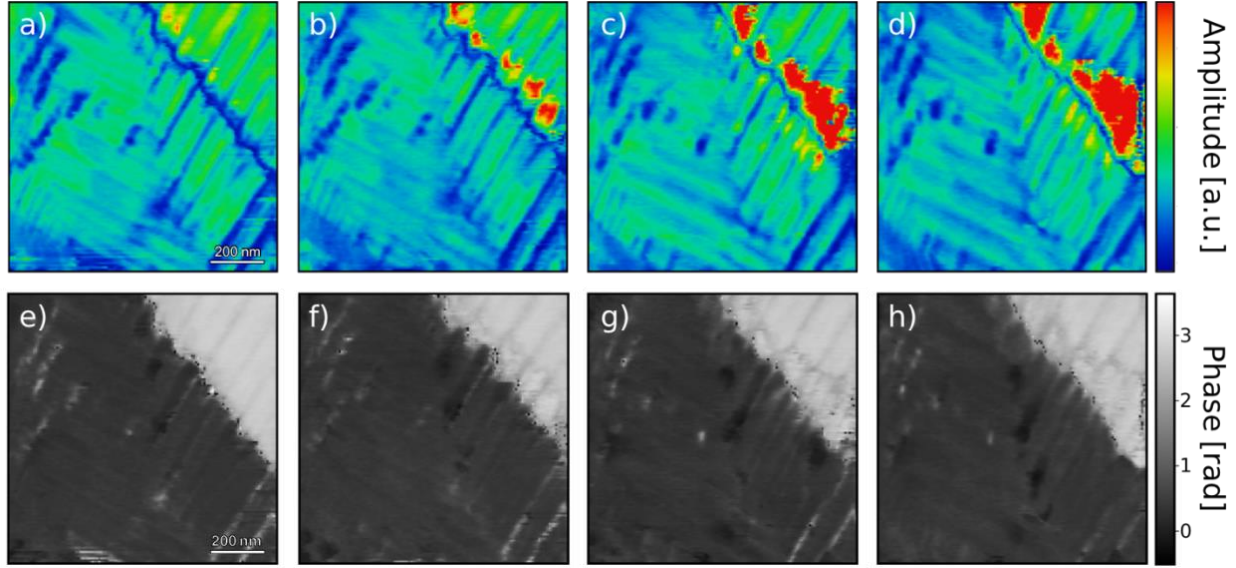

**Figure 3:** Single frequency PFM amplitude (a-d) and phase (e-h) of FerroBot applying +10V, 10ms pulse at the domain wall. Left to right columns are FerroBot cycles 1, 2, 7, and 13, respectively. FerroBot set to trigger under negative to positive phase switching; thus, domain is 'pushed' to top right. Note amplitude dependence on cycle illuminates polarization frustration on right side of domain wall.

Figure 3 shows a series of events engaging FerroBot manipulation of the domain wall between regions with orthogonal domain structures. Here, +10V bias pulses are applied at the 180° domain walls between the two regions with dense *a-c* domain structures, where the polarity of the *c*-domains, and correspondingly the in-plane direction of *a*-domains, changes. In this case, the application of the bias pulse exactly at the wall does not lead to macroscopic wall motion but changes the wall geometry on length scales comparable to the period of the *a-c* domain pattern. Notably, repeated processes lead to emergence of regions with enhanced electromechanical response, which are clearly visible in Fig. 3 (a- d).

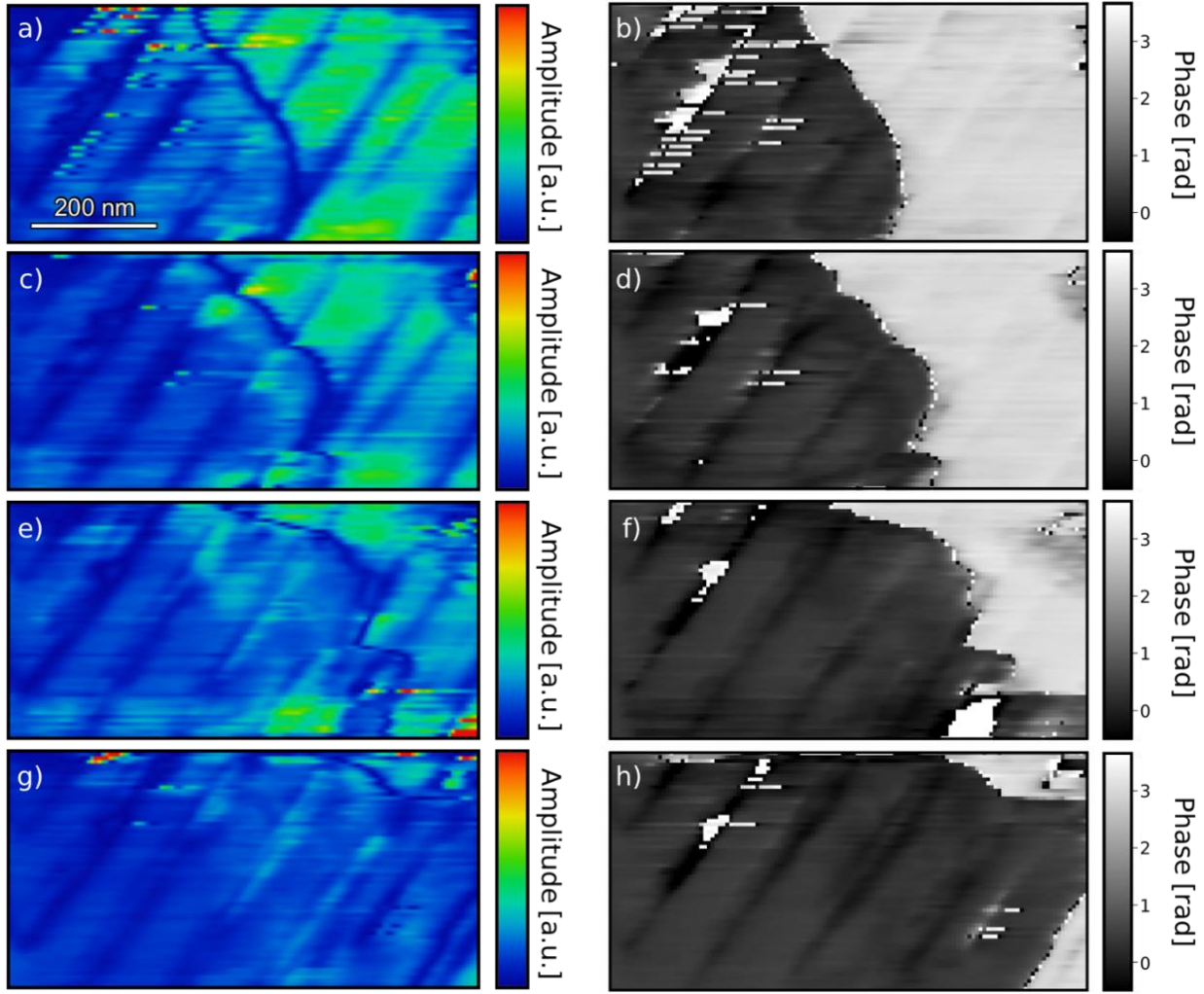

**Figure 4:** Single frequency PFM amplitude (left column) and phase (right column) of FerroBot applying +15V, 10ms pulse at the domain wall. Top to bottom rows are FerroBot cycles 1, 2, 3, and 4, respectively. Note this is same domain wall as shown in Figure 3. FerroBot set to trigger under positive to negative phase switching; thus, domain is 'pushed' to the right. In contrast to Figure 3, polarization is no longer frustrated as indicated by minimal/no amplitude increase in direction of domain wall motion.

In comparison (Fig. 4), the evolution of the PFM signal is shown for the case when the pulse magnitude is increased to +15V for the same domain wall. In this case, the bias is sufficient to unpin the domain wall, resulting in the mesoscopic wall motion. Note that in all cases the wall adopts a characteristic shape consistent with strong pinning of the 180° walls at the 90° walls oriented along the motion direction. In this case, the amplification of the electromechanical response at the driven wall is not observed.

To explain the observed behaviors, we argue that the 180° domain walls in the material with the dense *a-c* domain wall structure experience strong pinning at the 90° walls. During application of the subcritical bias values at the wall, it deforms into a local metastable free energy minimum corresponding to significant deviation of the polarization field from the crystallographically consistent ones, i.e., multiple regions with polarization rotation (Fig. 5). On the application of higher bias values, the domain wall moves macroscopic distances and is allowed to relax into lower-energy states, analogous to the difference between the static friction and dynamic friction states in mechanical systems. This behavior can be qualitatively understood from the response of a pinned domain wall to an applied force, both close to depinning and for small forces.[43,44] The application of a local force, even if smaller than the depinning force, can induce local motion of the wall and bring it to a local metastable state. Here, the SPM tip field is concentrated into a small volume, precluding the formation of an avalanche and effectively drives the system to a higher energy state (Fig. 3). A larger bias, even if local, can push the system to a state that will trigger an avalanche on a much larger length scale (Fig. 4). This process would thus be similar to what thermally activated processes induce in the creep regime,[45] for which the thermal nucleus can trigger a large avalanche, as confirmed numerically[44] and experimentally.[46] It would be interesting to see if such a local drive produces a similar distribution of avalanches than thermal processes. Another interesting aspect would be to correlate spatially such events as a function of the distance between two separate local drives. As a further relevant comparison, such high defect density states can be compared to the relaxor phase.

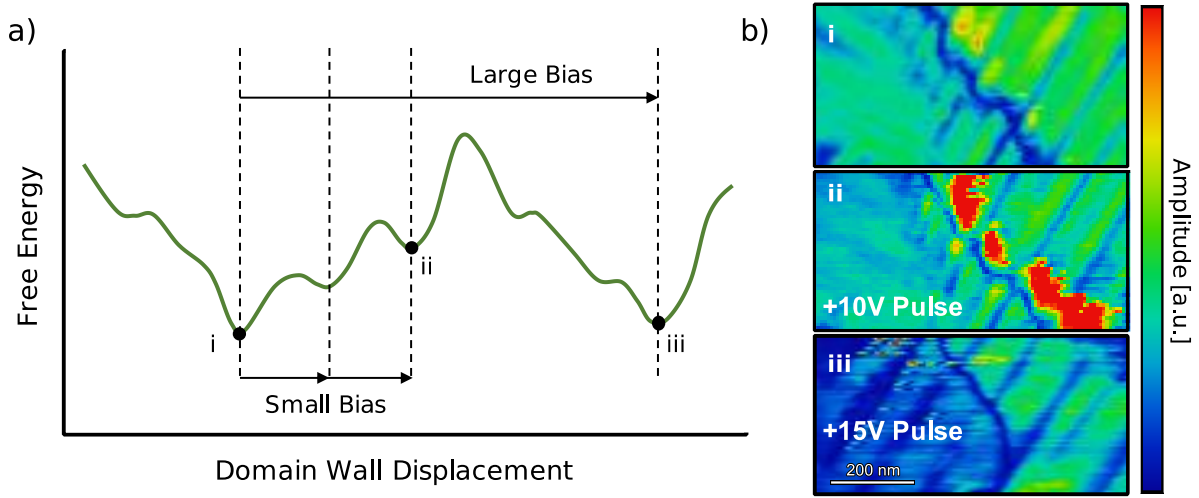

**Figure 5:** (a) Illustration of free energy landscape in a highly pinned ferroelectric system indicating small biases allow for metastable frustrated domain energies where large biases result in preferred lower energy minima. (b) Corresponding FerroBot images (from Figure 3a,c and 4a) for different free energy vs. domain wall displacement locations (i.e., i, ii, iii).

To understand the origin of the enhanced electromechanical responses observed in the non-equilibrium phase, we consider the general susceptibility of active ferroelectric materials. The mechanical deformation induced by the biased PFM probe in the ferroelectric film on a rigid substrate can be written via the corresponding elastic Green's function $G_{ij}^f(\mathbf{x}, \boldsymbol{\xi})$ for the second boundary value problem[47,48] with respect to displacement.[49] In the Fourier $\mathbf{k}$-domain, the surface displacement can be represented as:

$$\tilde{U}_i(\mathbf{q}, x_3 = 0) = \int_{-\infty}^{\infty} dk_1 \int_{-\infty}^{\infty} dk_2 \int_0^h d\xi_3 \tilde{G}_{ij,k}(q_1 - k_1, q_2 - k_2, \xi_3) S_{jk}(k_1, k_2, \xi_3, \omega). \quad (1a)$$

The Fourier image of the kernel $S_{jk}(k_1, k_2, \xi_3, \omega)$ is:

$$S_{jk}(\mathbf{k}, \omega) = d_{jkl}\tilde{P}_l(\mathbf{k}, \omega) + Q_{jklm}\tilde{P}_l(\mathbf{k}, \omega) * \tilde{P}_m(\mathbf{k}, \omega). \quad (1b)$$

Here, $\tilde{P}_l$ is the Fourier transform of ferroic polarization, $(\tilde{P}_l * \tilde{P}_m)$ is the Fourier convolution corresponding to the Fourier image of the product $P_l(\boldsymbol{\xi})P_m(\boldsymbol{\xi})$, $d_{jkl}$ is a piezoelectric tensor, which is nonzero for a low symmetry paraelectric phase, $Q_{mjkl}$ is the electrostriction strain tensor, and $\omega$ is the PFM tip excitation frequency. Since the components of the Green tensor have no poles except for the first order pole at **k**=0[47–50], it reflects the long-range nature of elastic fields without any specific "local" scale.

The polarization components obey nonlinear dynamic equations of Landau-Ginzburg-Devonshire (LGD) type.[50] The electric field $E_i(\mathbf{r},t)$ is the sum of the PFM tip field, $E_i^{ext}(\mathbf{r},t)$, and the internal depolarization field, $E_i^{dep}(\mathbf{r},t)$, which can be found from the corresponding Laplace's equation. Given that the probe field is much smaller than the coercive one, the Fourier image of linearized polarization response to the electric excitation by a PFM tip can be estimated as:

$$\widetilde{P}_i(\mathbf{k},\omega) \approx \widetilde{P}_i^S(\mathbf{k},\omega) + \frac{\widetilde{E}_i^{ext}(\mathbf{k},\omega) - d_{ijk}\widetilde{u}_{jk}(\mathbf{k},\omega)}{i\omega\Gamma + b_{ii}(T,\mathbf{k},\omega) + g_{iilm}k_l k_m}, \quad (2a)$$

where $\widetilde{P}_i^S(\mathbf{k},\omega)$ is the spontaneous polarization, $i$=1, 2, or 3 (without summation), $\Gamma$ is a kinetic coefficient, and $g_{iilm}$ are the gradient tensor component(s) in the time-dependent LGD equation. Functions $b_{ij}(T,\mathbf{k},\omega)$ are renormalized by local strains, $\widetilde{u}_{jk}(\mathbf{k},\omega)$, and local depolarization effects,

$$b_{ij}(T,\mathbf{k},\omega) = a_{ij}(T) + Q_{ijkl}\widetilde{u}_{kl}(\mathbf{k},\omega) + \frac{\eta_{ij}(\mathbf{k},\omega)}{\varepsilon_0\varepsilon_b} + a_{ijkl}\widetilde{P}_l^S(\mathbf{k},\omega) * \widetilde{P}_m^S(\mathbf{k},\omega). \quad (2b)$$

The coefficient, $a_{ij}(T)$, linearly depends on temperature and changes its sign at Curie temperature $T_C$. $\widetilde{u}_{jk}(\mathbf{k},\omega)$ are local strains, $\eta_{ij}(\mathbf{k},\omega)$ is an effective depolarization factor, and $a_{ijkl}$ are the components of the 3rd order polarization terms in the time-dependent LGD equation. Expressions in Eqs. (2) are derived in Supplemental Materials. Substitution of Eqs. (2) to Eqs. (1) and decoupling approximations lead to the following estimations of the kernel (1b):

$$S_{jk}^{FE}(\mathbf{k},\omega) \approx d_{jkl}\widetilde{P}_l^S(\mathbf{k},\omega) + Q_{jklm}\widetilde{P}_l^S * \widetilde{P}_l^S(\mathbf{k},\omega) + \frac{d_{jkl}\widetilde{E}_l^{ext}(\mathbf{k},\omega) + Q_{jklm}\widetilde{P}_l^S * \widetilde{E}_m^{ext}(\mathbf{k},\omega)}{i\omega\Gamma + b_{ll}^*(T,\mathbf{k},\omega) + g_{llpn}k_p k_n}. \quad (3)$$

In Eq. (3) we neglect one electrostrictive strain $\sim Q_{jklm}\widetilde{E}_l^{ext} * \widetilde{E}_m^{ext}(\mathbf{k},\omega)$ in comparison with local piezoelectric strains $\sim Q_{jklm}\widetilde{P}_l^S(\mathbf{k},\omega) - d_{lmn}\widetilde{u}_{mn}(\mathbf{k},\omega)$, as well as the terms $O[\widetilde{u}_{mn}^2(\mathbf{k},\omega)]$.

The maxima (or poles at $\omega\Gamma \to 0$) of denominator in Eq. (3) correspond to the maxima (or divergence at $\omega\Gamma \to 0$) of linear dynamic susceptibility spectra, $\chi_{ii}(\mathbf{k},\omega) \sim dP_i(\mathbf{k},\omega)/dE_i^{ext}(\mathbf{k},\omega)$. The susceptibility spectrum is proportional to:

$$\chi_{ii}^{FE}(\mathbf{k},\omega) \approx \frac{1}{i\omega\Gamma + b_{ii}^*(T,\mathbf{k},\omega) + g_{iilm}k_l k_m}. \quad (4)$$

The susceptibility is maximal (if $\omega\Gamma > 0$) or diverges (if $\omega\Gamma \to 0$) when $b_{ii}^*(T,\mathbf{k},\omega) + g_{iilm}k_l k_m \to 0$.

The inverse Fourier transform is impossible to derive in the general case. However, since both the elastic Green function and the probe electric field have no poles except for the first order pole at $\mathbf{k}=0$, the scale of the local PFM response is defined by the minima or zeros of the generalized susceptibility denominator. Correspondingly, the regions of maximal linear dynamic susceptibility in the $\mathbf{r}$-domain are the most probable candidates for the regions of maximal local PFM response. The first candidates for these points are in the immediate vicinity (about several correlation lengths $\frac{1}{R_{ii}^C} \cong \sqrt{\frac{1}{g_{ii}^{eff}}\left|-2a_{ii}(T) + Q_{iikl}u_{kl} + \frac{\eta_{ii}}{\varepsilon_0\varepsilon_b}\right|}$) of the curved domain walls with giant linear susceptibility, while the wall itself can give zero response from the symmetry theory.

While the analysis above provides the general mechanisms for the emergence of the enhanced electromechanical responses in the vicinity of disordered domain walls, this analysis is insufficient to make specific predictions (since at the ideal wall the response is zero due to symmetry). Thus, to explore the feasibility of the proposed mechanism we performed phase-field modeling of the domain wall dynamics and associated electromechanical responses (see Methods for details). We start with a PTO thin film consisting of two domain stripes: $c_+(001)/a_+(100)$ and $c_-(00\text{-}1)/a_-(\text{-}100)$, separated by a 180º domain wall (Fig. S3)). The PTO film is assumed to be fully relaxed on the KTiO3 substrate. When the film is in an equilibrium state, both 90º and 180º domain walls become slightly curved with a polarization frustration (Fig. 5(a)). Local tip biases along the (001) direction are then applied on the film surface at three different specific locations: (i) near the 90º domain wall, (ii) near the 180º domain wall, and (iii) inside the $c_+$ domain, as illustrated in Fig. 5(a); and the average out-of-plane strain ($\varepsilon_{33}$) beneath the probe is calculated with increasing bias. We reset the initial strains ($\varepsilon_{33}$) to be zero for all cases (i)-(iii) for easier comparison. From Fig. 5(c), the incremental out-of-plane strain ($\Delta\varepsilon_{33}$) is maximized when the tip is near the 90º and 180º domain walls, compared to inside the domain. This is further illustrated by the spatial profiles of $\varepsilon_{33}$ in the $x$-$z$ cross-section plane (Fig. 5b), in which a larger area of $\varepsilon_{33}$ enhancement is highlighted (red dashed circle) when the tip is near the 90º domain walls. It is also seen that the 90º domain wall slightly moves along $-x$ direction near the tip, indicating that the polarization rotates from the horizontal to vertical direction, which contributes to the local strain enhancement. The dependence of piezoelectric coefficients on the tip biases (calculated as $d_{33} = \Delta\varepsilon_{33}/\Delta E_3$) are compared for the three tip locations (Fig. 5(d)). We observe that the local piezoelectric response near the 90⁰ domain wall is almost three times as large as those inside the $c_+$ domain. We also moved the tip to different locations (Fig. S4(a)). In all cases, maximized $\varepsilon_{33}$ increment and $d_{33}$ are always found near the domain walls (Fig. S4(b),(c)). Our simulation indicates that a large polarization frustration, induced here by an applied tip voltage, at the domain walls causes higher piezoelectric responses.

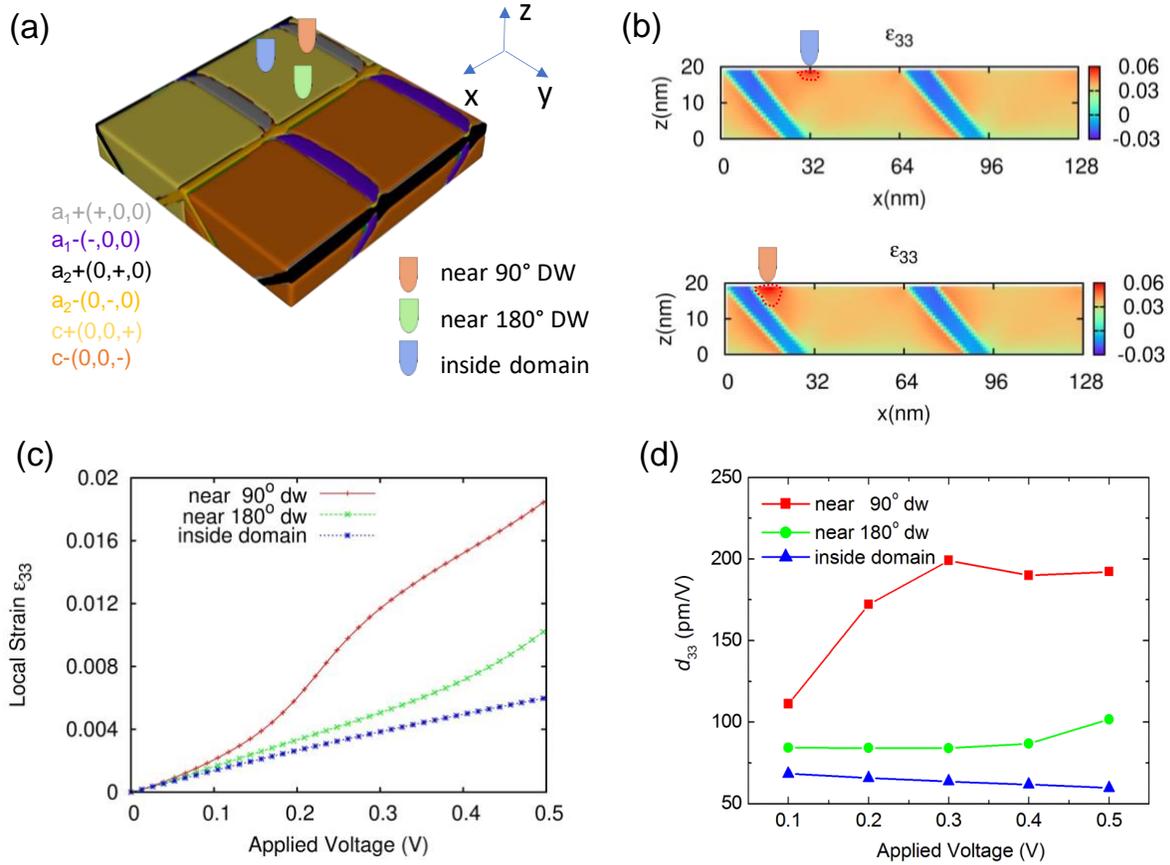

**Figure 6: Phase field modeling domain walls.** (a) 3D phase field model illustrating equilibrium domain structures of PTO. Colors represent different domains, with three probe locations: (i) near 90⁰ domain walls, (ii) near 180⁰ domain walls, and (iii) inside the $c_+$ domain. (b) Local strain ($\varepsilon_{33}$) distribution (in *x-z* cross-section plane) under tip bias of 1.3V, with red dashed circles represent areas of strain enhancement. (c) Local strain $\varepsilon_{33}$ (averaged in a $5 \times 5 \times 5$ nm³ volume beneath tip) with increasing tip bias in different locations. (d) Dependence of local piezoelectric coefficient ($d_{33}$) on tip bias calculated from average local strain in (c).

In summary, here we demonstrate a new paradigm for the domain and domain wall engineering in ferroelectric materials. Using active image-based feedback, we demonstrate the creation of artificial metastable states with enhanced electromechanical response. Interestingly, these states emerge at relatively small biases, which are sufficient to induce wall motion over small lengths but insufficient to unpin the wall though an avalanche-like process. This highly frustrated state possesses a higher electromechanical response due to enforced polarization frustration, potentially resembling that in ferroelectric relaxors. We pose that this approach is universal and the implementation of image-based feedback schemes allows a broad spectrum for experimental studies, including definition of local features, action tables, and use of non-local descriptors such as average domain structure and texture. Further developments are possible through incorporation of non-rectangular grids (as demonstrated by Ovchinnikov, et al.) and feedback-controlled paths.

This approach can in turn be employed to fabricate of the large sets of domains and topological defects, potentially scaling to macroscopic numbers, and exploring fundamental physics of ferroic walls, collective phenomena in ferroelectrics and relaxors, and low dimensional defects via automated experimentation. Further directions can include property-based feedback when the domain structure modification is performed until desired functionality is achieved.

**Acknowledgements:**

This work was supported (KK, RKV, NB) by the U.S. Department of Energy, Office of Science, Basic Energy Sciences, Materials Sciences and Engineering Division and performed at Oak Ridge National Laboratory's Center for Nanophase Materials Sciences, which is a US DOE Office of Science User Facility (SJ, SVK). And (TG) by the Swiss National Science foundation under Division II.

**Materials and methods:**

**Materials:**

The bilayer structure investigated via FerroBot consisted of 700nm polydomain PbTiO3 (PTO) grown on KTiO3 with SrRuO3 as a back electrode. Additional material details can be found at T. Nakashima *et al.* [51]

**Instrumentation:**

FerroBot measurements were employed via a modified ultra-high vacuum Omicron AFM-STM microscope controlled from a Nanonis real time controller framework. Signal processing and real time domain manipulation were driven by a coupled National Instruments USB-7856R multifunctional RIO and Stanford Research Systems RF lock-in amplifier (model SR844). Specifically, FerroBot was set to trigger under a negative to positive phase switch resulting in a user-defined voltage pulse being applied to the AFM tip. All experiments used Budget Sensor Multi75E-G Cr/Pt coated AFM probes (~3 N/m). All scans were taken at a scan speed of 0.8 um/s resulting in an applied bias (during FerroBot pulses) extending over a lateral length of 8 nm.

**Phase-field Simulation:**

In the phase-field simulation we employ the polarization vector $P_i = (P_x, P_y, P_z)$ as the order parameter to describe the ferroelectric state in the PbTiO3 thin film. The temporal evolution of $P_i$ is calculated by minimizing the total free energy $F$ with respect to $P_i$ via numerically solving the time-dependent Landau–Ginzburg–Devonshire (LGD) equations,

$$\frac{\partial P_i(\boldsymbol{x},t)}{\partial t} = -L \frac{\delta F}{\delta P_i(\boldsymbol{x},t)}, (i = 1 \sim 3) \qquad (5)$$

where $\boldsymbol{x}$ is the spatial position, $t$ is time, $L$ is the kinetic coefficient related to the domain-wall mobility. The total free energy $F$ of the PbTiO3 thin film includes the Landau, gradient, elastic, and electrostatic energies, which can be written as,

$$F = \int_V [f_{\text{lan}}(P_i) + f_{\text{grad}}(P_{i,j}) + f_{\text{elas}}(P_i, \varepsilon_{ij}) + f_{\text{elec}}(P_i, E_i)] dV , \qquad (6)$$

where $V$ is the total volume of the system, $\varepsilon_{ij}$ and $E_i$ denote the components of strain and electric fields. Detailed expressions of each free energy density can be found in the Reference [Huan Ferroelectrics 1987]. Equation (5) is numerically solved using a semi-implicit spectral method [L. Q. Chen and J. Shen, Comput. Phys. Commun. 108(2–3), 147–158 (1998).] based on a 3D geometry sampled on a $128\Delta x \times 128\Delta y \times 32\Delta z$ system size, with $\Delta x = \Delta y = \Delta z = 1.0$ nm. The thickness of the film, substrate, and air are $20\Delta z$, $10\Delta z$, and $2\Delta z$, respectively. The temperature is $T = 25°C$, and an isotropic relative dielectric constant ($\kappa_{ii}$) is chosen to be 50. The gradient energy coefficients are set to be $G_{11}/G_{110} = 0.6$, while $G_{110} = 1.73 \times 10^{-10} \text{C}^{-2}\text{m}^4\text{N}$. The substrate strain is set to be 0.5% due to the lattice mismatch between the PbTiO$_3$ thin film and the KTiO$_3$ substrate. The Landau coefficients, electrostrictive coefficients, and elastic-compliance constants are collected from Ref. [Huan Ferroelectrics 1987]

To model a local contact of a scanning probe tip, we specify the electrical potential distribution on the bottom ($\phi_1$) and top surface ($\phi_2$) of the PbTiO$_3$ film as,

$$\phi_1(x,y) = 0$$
$$\phi_2(x,y) = \phi_0 \frac{\gamma^2}{(x-x_0)^2 + (y-y_0)^2 + \gamma^2} \quad (7)$$

in which $\phi_0$ is the tip bias, $(x_0, y_0)$ is the tip position and $\gamma$ is the half width half magnitude (HWHM) of the tip. Here we chose $\gamma = 5$nm. The local strain ($\varepsilon_{33}$) and electric field (E$_3$) are averaged in a $5 \times 5 \times 5$ (nm$_3$) cube beneath the PFM probe. The applied voltages increase from 0.0V to 0.5V to scale with real experiments (10V on a 700nm thick film).

**Supplementary Information**

**A.1. Formulation of general problem**

The linear partial differential equation defining the mechanical displacement vector $U$ for a ferroic (paraelectric, relaxor, ferroelectric) film on a rigid substrate has the form:

$$\begin{cases} \dfrac{\partial \sigma_{ij}(\mathbf{x},t)}{\partial x_j} + \rho \dfrac{\partial^2 U_i}{\partial t^2}(\mathbf{x},t) = 0, \\ \sigma_{3i}(x_3=0)=0, \quad U_i(x_3=h)=0. \end{cases} \quad (A.1)$$

Here $\rho$ is the mass density, $\sigma_{ij}$ is a stress tensor. In the limit of semi-infinite ferroic the boundary condition $U_i(x_3=h)=0$ is substituted by the condition of stress absence. The time-dependent term $\rho \dfrac{\partial^2 U_i}{\partial t^2}$ is very small for excitation pulses at frequency $\omega$ much smaller than the characteristic frequencies of acoustic phonons, $\omega_a$. Hereinafter we will assume that $(\omega/\omega_a)^2 \ll 1$, since PFM frequency ranges from mHz to MHz.

Assuming that the electromechanical coupling in a considered ferroic is described by an arbitrary tensor of piezoelectric and / or electrostrictive strains, generalized Hooke's law takes the form

$$\sigma_{ij} = c_{ijkl} u_{kl} - d_{ijk} P_k - Q_{ijkl} P_k P_l, \quad (A.2)$$

Where $c_{ijkl}$ is tensor of elastic stiffness; $u_{kl} = \dfrac{1}{2}\left(\dfrac{\partial U_k}{\partial x_l} + \dfrac{\partial U_l}{\partial x_k}\right)$ is elastic strain tensor; $d_{mjk}$ is a piezoelectric tensor, that is nonzero for a low symmetry paraelectric phase. Otherwise effective piezoelectric response appears in the ferroelectric phase only, as linearized electrostriction. $P_i(\mathbf{r},t)$ is a ferroic polarization. The electrostriction strain tensor $Q_{mjkl}$ can be renormalized by the Maxwell stresses.

The solution of the stationary equation (A.1) with boundary conditions, $\sigma_{3i}(x_3=0)=0$, written through the corresponding elastic Green's function $G_{ij}^f(\mathbf{x},\xi)$ for the second boundary value problem [i, ii] with respect to displacements is [iii]:

$$U_i(\mathbf{x}) = \int_{\xi_3=0}^{h}\int_{\xi_2=-\infty}^{\infty}\int_{\xi_1=-\infty}^{\infty} \dfrac{\partial G_{ij}^f(\mathbf{x},\xi)}{\partial \xi_k}\left[d_{jkl}P_l(\xi,t) + Q_{jklm}P_l(\xi,t)P_m(\xi,t)\right]d\xi_1 d\xi_2 d\xi_3. \quad (A.3)$$

Here $h$ is the film thickness.

Since the components of $G_{ij}^f(\mathbf{x},\xi)$ in (A.3) depend on the coordinate difference $x_1 - \xi_1$ and $x_2 - \xi_2$ [iv], the 2D Fourier transform (A.2) along the transverse coordinates $\{x_1, x_2\}$ can be written as:

$$\tilde{U}_i(\mathbf{q}, x_3 = 0) = \int_{-\infty}^{\infty} dk_1 \int_{-\infty}^{\infty} dk_2 \int_0^h d\xi_3 \tilde{G}_{ij,k}(q_1 - k_1, q_2 - k_2, \xi_3) \times \\ \left[d_{jkl}\tilde{P}_l(k_1, k_2, \xi_3) + Q_{jklm}(\tilde{P}_l * \tilde{P}_m)(k_1, k_2, \xi_3)\right] \quad \text{(A.4a)}$$

Here $(\tilde{P}_l * \tilde{P}_m)$ is the Fourier convolution corresponding to the Fourier image of the product $P_l(\xi)P_m(\xi)$. Green function and its derivatives are

$$G_{ij}(x_1, x_2, x_3 = 0, \xi) = \frac{1}{2\pi}\int_{-\infty}^{\infty} dk_1 \int_{-\infty}^{\infty} dk_2 \exp[-ik_1(x_1 - \xi_1) - ik_2(x_2 - \xi_2)]\tilde{G}_{ij}(k_1, k_2, \xi_3). \quad \text{(A.4b)}$$

$$\tilde{G}_{ij,l}(k_1, k_2, \xi) \equiv \begin{cases} ik_l \tilde{G}_{ij}(k_1, k_2, \xi), & l = 1,2 \\ \dfrac{\partial}{\partial \xi}\tilde{G}_{ij}(k_1, k_2, \xi), & l = 3 \end{cases} \quad \text{(A.4c)}$$

Since that the components of Green tensor have no poles, except for the first order pole at $\mathbf{k}=0$, e.g. $\tilde{G}_{ij}(k_1, k_2, \xi) \sim \dfrac{\exp(-\xi k)}{k}\left(A_{ij} + B\dfrac{k_i k_j}{k^2} + ...\right)$ [i-iv], it reflects the long-range nature of elastic fields without any specific "local" scale.

Polarization components obey nonlinear dynamic equation of e.g. LGD type [v]:

$$\Gamma \frac{\partial P_i(\mathbf{r},t)}{\partial t} + a_{ij}(T)P_j + a_{ijkl}P_j P_k P_l + ... - g_{ijkl}\frac{\partial^2 P_j}{\partial x_k \partial x_l} + (d_{ijk} + Q_{ijkl}P_l)u_{jk}(\mathbf{r},t) = E_i(\mathbf{r},t). \quad \text{(A.5a)}$$

The coefficient $a_{ij}(T)$ linearly depends on temperature and changes its sign at Curie temperature $T_C$. The boundary conditions to Eq.(A.5a) are of the third kind and accounts for the flexoelectric effect:

$$\left(g_{ijkl}\frac{\partial P_k}{\partial x_l} - F_{klij}\sigma_{kl}\right)n_j\bigg|_{x_3=h} = 0 \quad \text{(A.5b)}$$

where $\mathbf{n}$ is the outer normal to the film surfaces, $F_{klij}$ is a flexoelectric effect tensor.

Electric field $E_i(\mathbf{r},t)$ is the sum of PFM tip field, $E_i^{ext}(\mathbf{r},t)$, and internal depolarization field, $E_i^{dep}(\mathbf{r},t)$. Since the excitation frequency $\omega$ of SPM tip does not exceed MHz range, the field should be found from electrostatic equations. They are Laplace equation for electrostatic potential $\varphi$ outside ferroic, and Poisson equation inside it:

$$\varepsilon_0 \varepsilon_e \left( \frac{\partial^2}{\partial x_1^2} + \frac{\partial^2}{\partial x_2^2} + \frac{\partial^2}{\partial x_3^2} \right) \varphi = 0, \quad -\infty \leq x_3 \leq 0, \tag{A.6a}$$

$$\varepsilon_0 \varepsilon_b \left( \frac{\partial^2}{\partial x_1^2} + \frac{\partial^2}{\partial x_2^2} + \frac{\partial^2}{\partial x_3^2} \right) \varphi = \frac{\partial P_i}{\partial x_i}, \quad 0 \leq x_3 \leq h. \tag{A.6b}$$

Electric field components $E_i$ are related to the electric potential $\varphi$ in a conventional way, $E_i = -\partial \varphi / \partial x_i$. Boundary conditions are: fixed potential at the surface of conducting tip, $\varphi(\mathbf{r} \in tip, t) = U(t)$, and grounded bottom electrode $\varphi(x_3 = h, t) = 0$ (or its vanishing at infinity for a semi-infinite sample); and electric displacement continuity at the tip-ferroic and ferroic-air contacts.

General solution of Eq.(A.6) can be represented in Fourier domain as

$$\tilde{\varphi}(\mathbf{k},\omega) = \tilde{\varphi}_{dep}(\mathbf{k},\omega) + \tilde{\varphi}_{ext}(\mathbf{k},\omega), \qquad \tilde{\varphi}_{dep}(\mathbf{k},\omega) = \frac{1}{\varepsilon_0 \varepsilon_b k^2} \left[ k_i \tilde{P}_i(\mathbf{k},\omega) - L[k_i, \tilde{P}_i(\mathbf{k},\omega)] \right]. \tag{A.6c}$$

Here $k^2 = \sqrt{k_1^2 + k_2^2 + k_3^2}$. Since the original $P_i(\mathbf{r},t)$ is nonzero inside the film ($0 \leq x_3 \leq h$), the external potential $\tilde{\varphi}_{ext}(\mathbf{r},t)$ satisfies the boundary condition at the tip $\varphi_{ext}(\mathbf{r} \in tip, t) = U(t)$, and grounded bottom electrode or infinity. For the most common model of effective point charge for the probe field [i-iv] tells us that $\tilde{\varphi}_{ext}(\mathbf{k},\omega)$ has no poles, except for the first order pole at $\mathbf{k}=0$, e.g. $\tilde{\varphi}_{ext}(k_1,k_2,\xi_3) \sim \frac{\exp(-kd - k\xi_3/\gamma)}{4\pi\varepsilon_0 k}$ [i-iv], that reflects the long-range nature of quasi-static electric fields without any specific "local" scale.

The compensating "potential" $\tilde{\varphi}_c(\mathbf{k},\omega) = -\frac{L[k_i, \tilde{P}_i(\mathbf{k},\omega)]}{\varepsilon_0 \varepsilon_b k^2}$ provides zero boundary conditions for $\tilde{\varphi}_{dep}(\mathbf{r},t)$ and displacement continuity at the tip-ferroic and ferroic-air contacts. Mathematically $\Delta\varphi_c(\mathbf{r},t)$ is a linear integral function of polarization that satisfies Laplace equation $\Delta\varphi_c(\mathbf{r},t)=0$. In several simplest cases [vi] the depolarization field can be somehow described by a "local" depolarization factor, $\eta_{ij}(\mathbf{k},\omega)$, i.e.

$$\tilde{E}_i(\mathbf{k},\omega)\approx \tilde{E}_i^{ext}(\mathbf{k},\omega)-\frac{\eta_{ij}(\mathbf{k},\omega)}{\varepsilon_0\varepsilon_b}\tilde{P}_j(\mathbf{k},\omega). \quad (A.6d)$$

The formulation of the coupled problem (A.1)-(A.6) is too uncertain and complex for its solution in a general case of arbitrary domain structure. Below we discuss several approximate cases and make back-on-the envelope estimates.

### A.2. Frequency limits of polarization dynamics for different ferroics

The quasi-stationary solution of Eq.(A.5) for the polarization component $P_i(\mathbf{r},t)$ can be found in an adiabatic approximation, applicable when the corresponding Landau-Khalatnikov time $\tau_K = \Gamma/|a_{ii}|$ is much smaller than the characteristic times of external electric field changes, i.e. the strong inequality $\tau_K\omega \ll 1$ should be fulfilled. Only is this case one can regard that the polarization changes immediately follow the applied voltage pulses. For a proper ferroic far from the ferroelectric Curie point (e.g. at room temperature), $|a_{ii}|\cong(10^7-10^8)$ Jm/C$_2$, and so Landau-Khalatnikov time can be estimated as $\tau_K\cong(10^{-6}-10^{-9})$s. Thus the low-frequency limit means that $\omega \ll (10^6-10^9)$1/s. The estimates of $\tau_K$ for quantum paraelectrics with high dielectric susceptibility and especially ferroelectric relaxors with continuous spectra of relaxation times, can lead to much higher values, $\tau_K > (10^{-5}-1)$s and so it must be $\omega \ll (10^3-1)$1/s. That say we should not regard that the inequality $\tau_K\omega \ll 1$ is valid a priory.

### A.3. Effective polar and PFM responses

Regarding that the probe field is much smaller than the coercive one, the Fourier image of linear polarization response to the electric excitation by a PFM tip can be roughly estimated for paraelectric type (PE) response, as:

$$\tilde{P}_i(\mathbf{k},\omega)\approx \frac{\tilde{E}_i^{ext}(\mathbf{k},\omega)-d_{ijk}\tilde{u}_{jk}(\mathbf{k},\omega)}{i\omega\Gamma + a_{ii}^*(T,\mathbf{k},\omega)+g_{iilm}k_lk_m}, \quad (A.7a)$$

For a ferroelectric (FE) type response

$$\widetilde{P}_i(\mathbf{k},\omega) \approx \widetilde{P}_i^S(\mathbf{k},\omega) + \frac{\widetilde{E}_i^{ext}(\mathbf{k},\omega) - d_{ijk}\widetilde{u}_{jk}(\mathbf{k},\omega)}{i\omega\Gamma + b_{ii}^*(T,\mathbf{k},\omega) + g_{iilm}k_l k_m}. \tag{A.7b}$$

Where $i=1, 2$ or $3$ (without summation). Functions $a_{ij}^*(T,\mathbf{k},\omega)$ and $b_{ij}^*(T,\mathbf{k},\omega)$ are renormalized by e.g. local or misfit strains, $\widetilde{u}_{jk}(\mathbf{k},\omega)$, and local depolarization effects as

$$a_{ij}^*(T,\mathbf{k},\omega) = a_{ij}(T) + Q_{ijkl}\widetilde{u}_{kl}(\mathbf{k},\omega) + \frac{\eta_{ij}(\mathbf{k},\omega)}{\varepsilon_0 \varepsilon_b}. \tag{A.7c}$$

The coefficient

$$b_{ij}^*(T,\mathbf{k},\omega) = a_{ij}(T) + Q_{ijkl}\widetilde{u}_{kl}(\mathbf{k},\omega) + \frac{\eta_{ij}(\mathbf{k},\omega)}{\varepsilon_0 \varepsilon_b} + a_{ijkl}\widetilde{P}_l^S(\mathbf{k},\omega) * \widetilde{P}_m^S(\mathbf{k},\omega). \tag{A.7d}$$

Note that for a constant homogeneous spontaneous polarization $b_{ij}^*(T) \cong -2a_{ij}^*(T)$. Alternatively, the priory unknown convolution $\left(\widetilde{P}_j^S * \widetilde{P}_k^S\right)$ determines the coefficient $b_{ij}^*(T,\mathbf{k},\omega)$ in the presence of domain structure.

Substitution of Eqs.(A.7) to Eq.(A.4a) leads to the following estimations of the kernel

$$S_{jk}(\mathbf{k},\omega) = d_{jkl}\widetilde{P}_l(\mathbf{k},\omega) + Q_{jklm}\widetilde{P}_l(\mathbf{k},\omega) * \widetilde{P}_m(\mathbf{k},\omega) \tag{A.8a}$$

Under the absence of spontaneous polarization, the necessary condition for this is $a_{ii}^*(T) > 0$, the kernel is proportional to

$$S_{jk}^{PE}(\mathbf{k},\omega) \sim \frac{d_{jkl}\left[\widetilde{E}_l^{ext}(\mathbf{k},\omega) - d_{lmn}\widetilde{u}_{mn}(\mathbf{k},\omega)\right]}{i\omega\Gamma + a_{ll}^*(T,\mathbf{k},\omega) + g_{llmn}k_m k_n} + O\left[\widetilde{u}_{mn}^2(\mathbf{k},\omega)\right]$$
$$\frac{Q_{jklm}\left[\widetilde{E}_l^{ext}(\mathbf{k},\omega) - d_{lmn}\widetilde{u}_{mn}(\mathbf{k},\omega)\right] * \left[\widetilde{E}_m^{ext}(\mathbf{k},\omega) - d_{mqn}\widetilde{u}_{qn}(\mathbf{k},\omega)\right]}{\left(i\omega\Gamma + a_{ll}^*(T,\mathbf{k},\omega) + g_{llpn}k_n k_p\right)\left(i\omega\Gamma + a_{mm}^*(T,\mathbf{k},\omega) + g_{mmpn}k_n k_p\right)}, \tag{A.8b}$$

In decoupling approximation one can neglect the quadratic piezoelectric terms proportional to $d_{lmn}d_{jkl}$ as well as the terms proportional to $Q_{jklm}d_{lmn}$.

Using decoupling approximation for the case when spontaneous polarization $\widetilde{P}_l^S(\mathbf{k},\omega)$ is present. Under the necessary condition $b_{ii}^*(T) < 0$ the kernel is proportional to

$$S_{jk}^{FE}(\mathbf{k},\omega) \sim d_{jkl}\tilde{P}_l^S(\mathbf{k},\omega) + \frac{d_{jkl}\tilde{E}_l^{ext}(\mathbf{k},\omega)}{i\omega\Gamma + b_{ll}^*(T,\mathbf{k},\omega) + g_{llmn}k_m k_n} + Q_{jklm}\tilde{P}_l^S * \tilde{P}_m^S(\mathbf{k},\omega)$$

$$+ \frac{Q_{jklm}\tilde{P}_l^S(\mathbf{k},\omega) * \tilde{E}_m^{ext}(\mathbf{k},\omega)}{i\omega\Gamma + b_{ll}^*(T,\mathbf{k},\omega) + g_{llpn}k_n k_p} + \frac{Q_{jklm}\tilde{E}_l^{ext}(\mathbf{k},\omega) * \tilde{E}_m^{ext}(\mathbf{k},\omega)}{\left(i\omega\Gamma + b_{ll}^*(T,\mathbf{k},\omega) + g_{llpn}k_n k_p\right)\left(i\omega\Gamma + b_{mm}^*(T,\mathbf{k},\omega) + g_{mmpn}k_n k_p\right)}$$

, (A.8c)

The numerators in Eqs.(A.8) are proportional to the ***probe field*** ($\tilde{E}_l^{ext}(\mathbf{k},\omega)$, ***local piezoelectric*** ($\sim Q_{jklm}\tilde{P}_l^S(\mathbf{k},\omega) - d_{lmn}\tilde{u}_{mn}(\mathbf{k},\omega)$)) and ***electrostrictive*** ($\sim Q_{jklm}\tilde{E}_l^{ext} * \tilde{E}_m^{ext}(\mathbf{k},\omega)$)) strains. Typically one can neglect the terms $O[\tilde{u}_{mn}^2(\mathbf{k},\omega)]$.

The maxima (poles at $\omega\Gamma \to 0$) of denominator in Eqs.(A.8b)-(A.8c) correspond to the maxima (or divergence at $\omega\Gamma \to 0$) of linear dynamic susceptibility spectra, $\chi_{ii}(\mathbf{k},\omega) \sim dP_i(\mathbf{k},\omega)/dE_i^{ext}(\mathbf{k},\omega)$. The susceptibility spectra is proportional to

$$\chi_{ii}^{PE}(\mathbf{k},\omega) \approx \frac{1}{i\omega\Gamma + a_{ii}^*(T,\mathbf{k},\omega) + g_{iilm}k_l k_m}, \qquad a_{ii}^*(T) > 0, \qquad (A.9a)$$

$$\chi_{ii}^{FE}(\mathbf{k},\omega) \approx \frac{1}{i\omega\Gamma + b_{ii}^*(T,\mathbf{k},\omega) + g_{iilm}k_l k_m}. \qquad b_{ii}^*(T) < 0. \quad (A.9b)$$

Susceptibilities are maximal (if $\omega\Gamma > 0$) or diverges (if $\omega\Gamma \to 0$) when $a_{ii}^*(T,\mathbf{k},\omega) + g_{iilm}k_l k_m \to 0$ or $b_{ii}^*(T,\mathbf{k},\omega) + g_{iilm}k_l k_m \to 0$. The conditions are written in **k**-domain, and the functions included here are defined by Eqs.(A.7c-d). Inverse Fourier transform is impossible to do in general case. However, since both elastic Green function and probe electric field have no poles except for the first order pole at $k=0$, the scale of local PFM response is indeed defined by the minima or zeros of generalized susceptibility denominator. Notably that the susceptibility is, in tightly related with polarization fluctuations correlation function [vii, viii]. Corresponding (anisotropic) correlation lengths $R_{ij}^C$ can be roughly estimated as:

$$\frac{1}{R_{ii}^C} \cong \sqrt{\frac{1}{g_{ii}^{eff}}\left|a_{ii}(T) + Q_{iikl}u_{jk} + \frac{\eta_{ii}}{\varepsilon_0 \varepsilon_b}\right|}, \quad a_{ii}(T) > 0, \qquad (A.10a)$$

$$\frac{1}{R_{ii}^C} \cong \sqrt{\frac{1}{g_{ii}^{eff}}\left|-2a_{ii}(T) + Q_{iikl}u_{kl} + \frac{\eta_{ii}}{\varepsilon_0 \varepsilon_b}\right|}, \quad a_{ii}(T) < 0. \qquad (A.10b)$$

Where $g_{ii}^{eff}$ is a positive combination of gradient coefficients $g_{iikk}$. Numerical estimates of the correlation lengths strongly depend on temperature, local strains and depolarization factors. Typically it ranges from nm to hundreds of nm, and apparently diverges when the system approaches the second order phase transition to a paraelectric phase.

Being mathematically very uncertain due to the aforementioned impossibility to make the inverse Fourier transform in general case, Eqs.(A.9)-(A.10) indicate on a well-known fact, that the points of maximal linear dynamic susceptibility in **r**-domain are the most probable candidates for the regions of maximal local PFM response (see **Figs S1-S2**). The first candidates for these points are the immediate vicinity (about several $R_{ii}^C$) of curved domain walls with giant linear susceptibility, while the walls itself can give zero response from the symmetry theory.

Actually, the region of negative real part of susceptibility, $Re[\chi_{ii}^{FE}(\boldsymbol{k},\omega)]$, exists for $-1<kR_c<1$ corresponding to the nominally uncharged 180-degree domain wall and dimensionless frequencies ω=0.0005 – 0.1 (in $\tau_K$ units) being relatively low in comparison with phonon relaxation time $\tau_K \approx \frac{\Gamma}{|a_{11}|}$. [see **Fig.S1(a)**]. Exactly the region $Re[\chi_{ii}^{FE}(\boldsymbol{k},\omega)] < 0$ corresponds to nonzero imaginary part, $Im[\chi_{ii}^{FE}(\boldsymbol{k},\omega)]$ [see **Fig.S1(b)**]. Corresponding gain factor $-\frac{Re[\chi_{ii}^{FE}(\boldsymbol{k},\omega)]}{|Im[\chi_{ii}^{FE}(\boldsymbol{k},\omega)]|}$, which is responsible for the amplification of PFM response signal, is positive in the above region and reaches high values (~5 – 500) for dimensionless frequencies within the range ω=(0.0005 – 0.1) [see solid, dashed and dotted black curves in **Fig.S1(c)**].

The inverse Fourier transform of real and imaginary parts of $S(\boldsymbol{k},\omega)$, shown by black curves in **Fig.S2(a,b)**, returns the effective piezoelectric response $PR(\boldsymbol{x},\omega)$, which is zero at the wall and has maxima or oscillate in the wall vicinity at different frequencies ω=(0.0005 – 0.1) [see black curves in **Fig.S2(c,d)**]**.** X-profile of the nominally uncharged 180-degree domain wall is shown by red curve in **Fig.S2 (c,d)** for comparison.

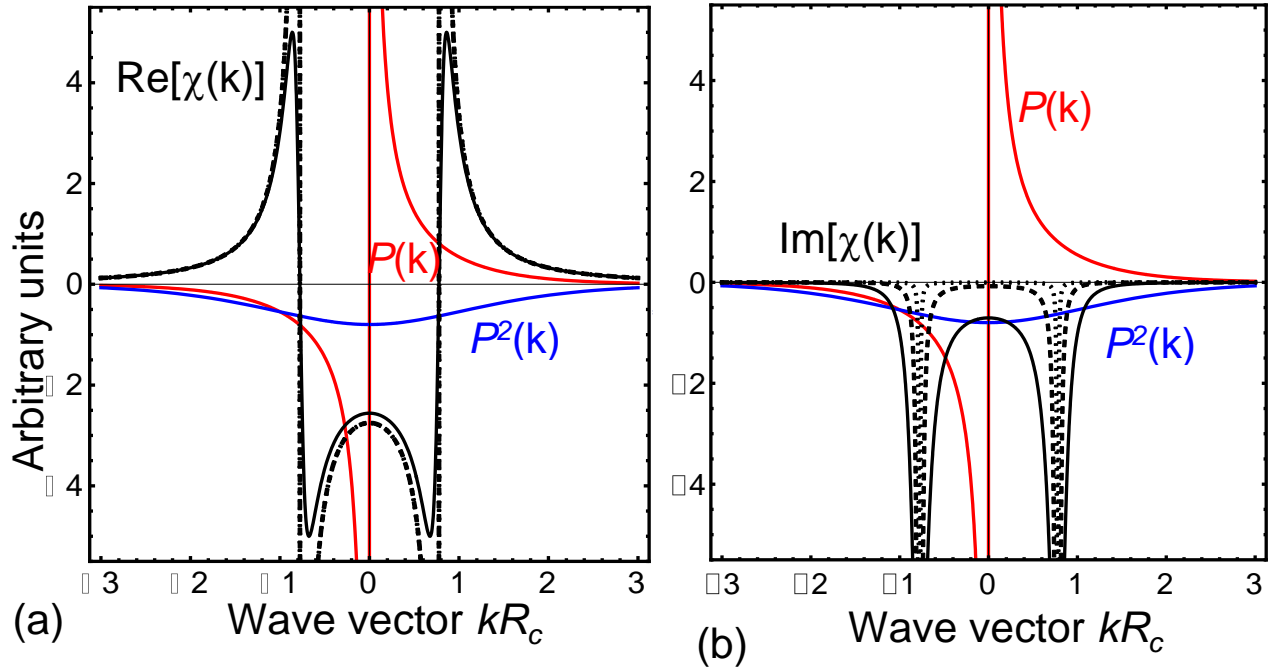
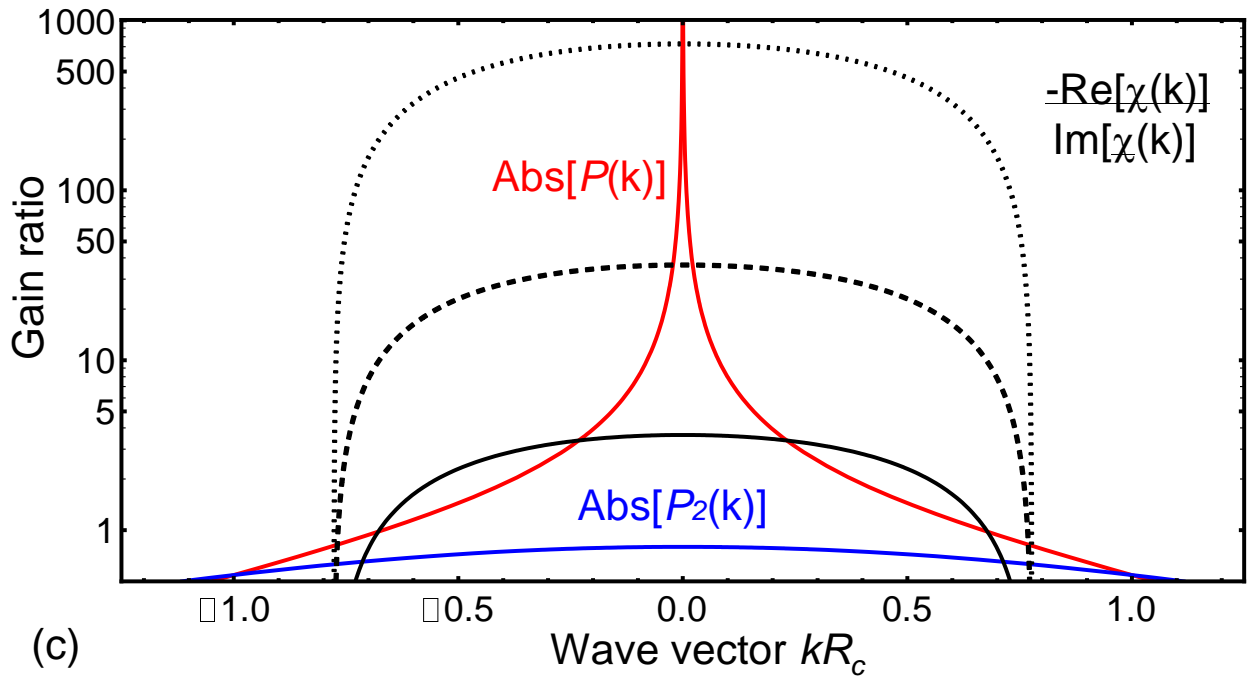

**Figure S1.** **(a, b)** The Fourier transform of polarization across a nominally uncharged 180-degree domain wall (red curves), convolution $\tilde{P}_l(\mathbf{k}, \omega) * \tilde{P}_m(\mathbf{k}, \omega)$ (blue curves), real **(a)** and imaginary **(b)** parts of generalized susceptibility $\chi_{ii}^{FE}(\mathbf{k},\omega)$, and gain factor $-\frac{Re[\chi_{ii}^{FE}(\mathbf{k},\omega)]}{|Im[\chi_{ii}^{FE}(\mathbf{k},\omega)]|}$ (c) calculated at 3 different frequencies (black curves). Black curves are calculated at dimensionless frequencies $\omega_1=5\cdot10^{-4}$ (dotted), $\omega_2=0.01$ (dashed) and $\omega_3=0.1$ (solid).

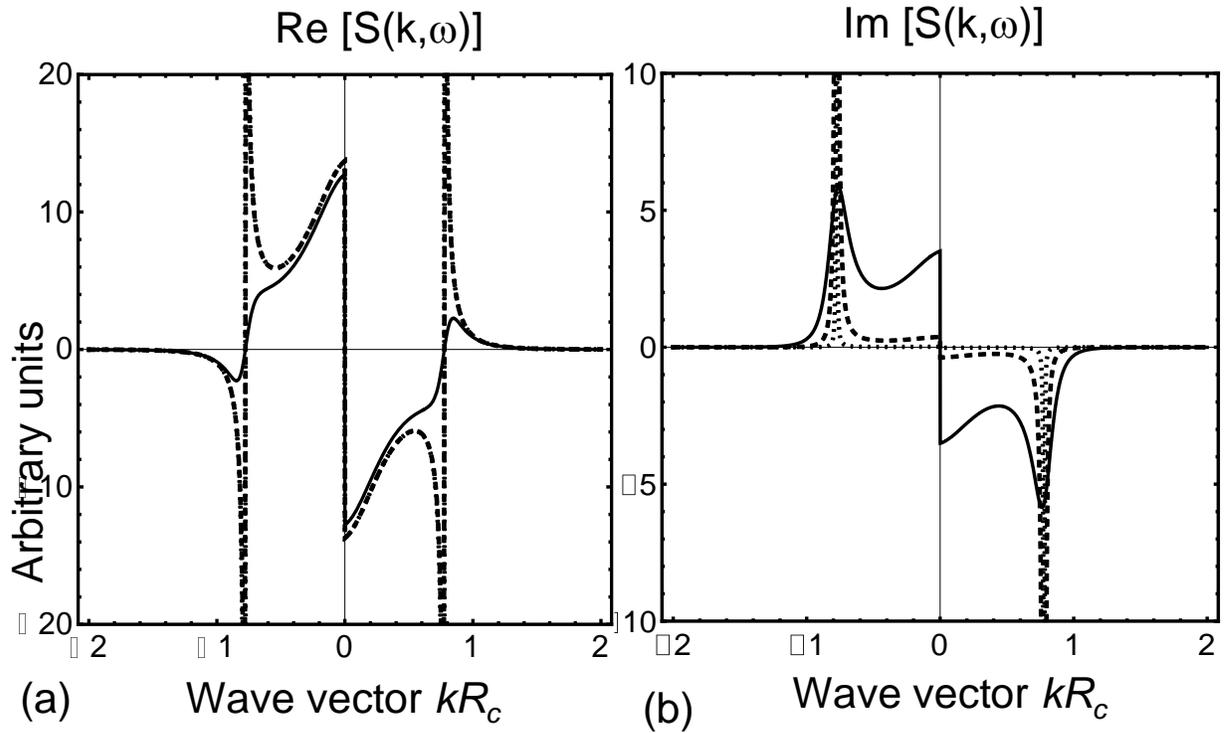

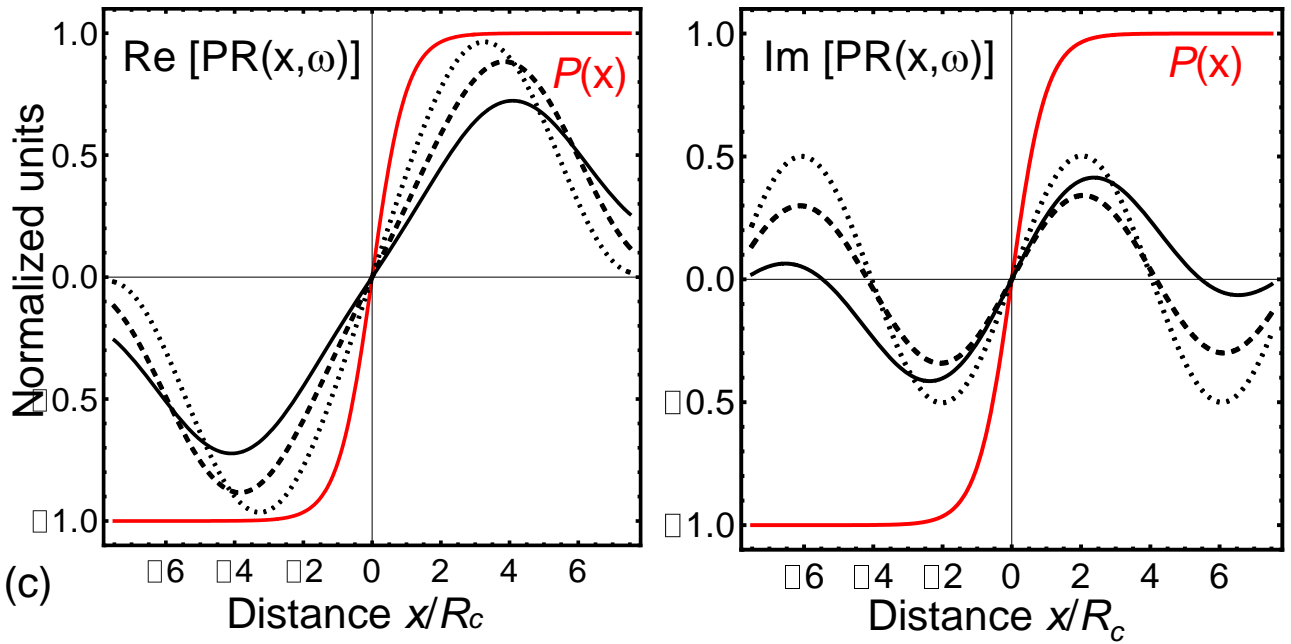

**Figure S2.** The Fourier transform of real (**a**) and imaginary (**b**) parts of Green function kernel, $S(\mathbf{k}, \omega)$, for the model case of polarization profile across a nominally uncharged 180-degree domain wall, shown by red curves in **Fig. S1**, calculated at 3 different frequencies (black curves). The Fourier transform of real (**c**) and imaginary **(d)** parts of effective piezoelectric response (PR) calculated at 3 different frequencies (black curves)**.** Red curve in **(c,d)** is an x-profile of the nominally uncharged 180-degree domain wall. Black curves are calculated at dimensionless frequencies $\omega_1 = 5 \cdot 10^{-4}$ (dotted), $\omega_2 = 0.01$ (dashed) and $\omega_3 = 0.1$ (solid).

## B.1. Phase-field simulation of piezoelectric response

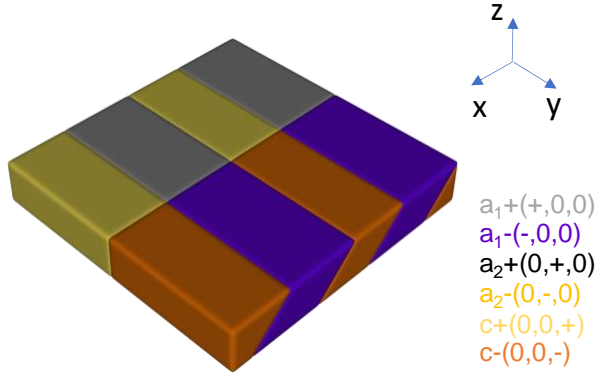

**Figure S3:** The initial present domain structure of PbTiO$_3$ thin film in the phase-field simulation

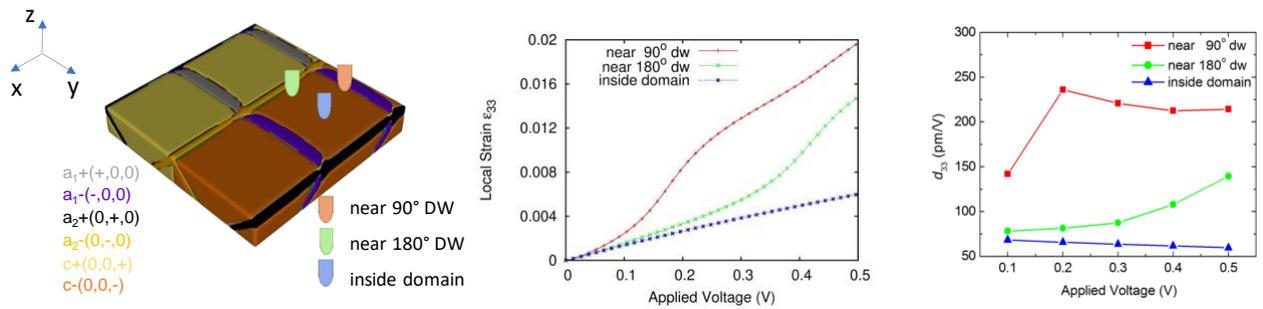

**Figure S4:** Phase-field simulation of PTO thin film under 0.0%. (a) Equilibrium domain structure; (b) The dependence of local strain beneath the top on the applied voltages; (c) The dependence of $d_{33}$ on applied voltages for different tip locations.

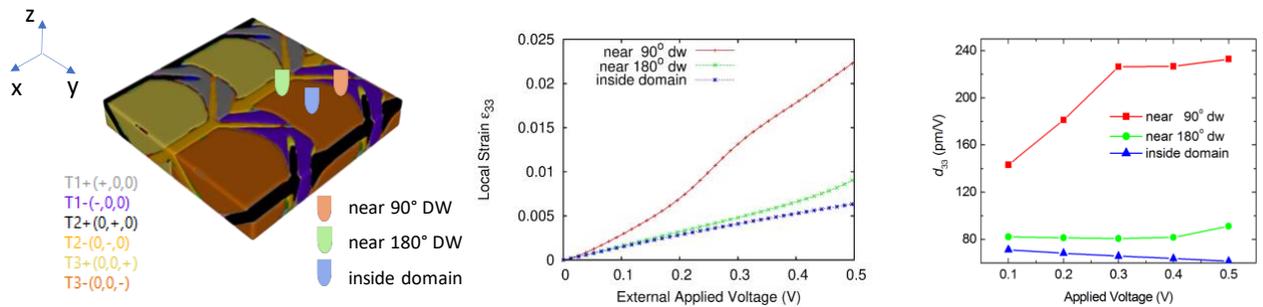

**Figure S5:** Phase-field simulation of PTO thin film under 0.5%. (a) Equilibrium domain structure; (b) The dependence of local strain beneath the top on the applied voltages; (c) The dependence of $d_{33}$ on applied voltages for different tip locations. Similar trends are observed as compared to 0.0% strained case.

**C.1 Reproduced polarization frustration with high stiffness cantilever.**

Cantilever dynamics can heavily influence the results obtained in piezoresponse force microscopy (PFM); thus, the following is a subset of results reproducing the experiment seen Figure 3 in efforts to eliminate the spurious electrostatics contribution to the PFM signal and purely measure electromechanical response. As such, the following results were obtained using a metal coated Nanosensor AFM probe with a high spring constant, ~40 N/m.

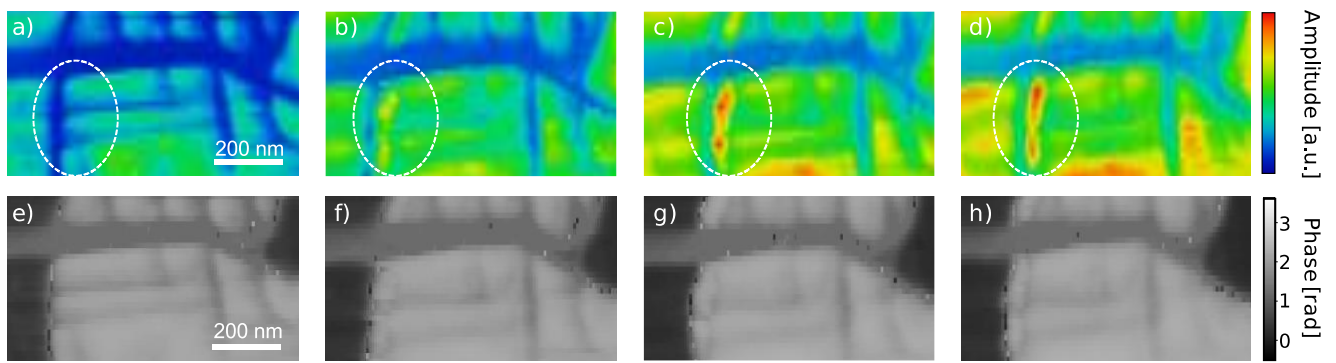

**Figure S6:** Single frequency PFM images (PFM amplitude (a-d) and PFM phase (e-h)) of ferroelectric domain frustration achieved using 5V, 10ms pulses at FerroBot cycles 0 (a,e), 1(b,f), 2(c,g), and 3(d,h).

As seen in Figure S6, the domain frustration and consequently enhanced electromechanical response is clearly observed (white dotted ellipse). Note, the overall amplitude in Figure 6(a-d) increases in addition to the frustrated domain, which we attribute to a resonance frequency shift. Nevertheless, the enhancement is still observed.